\documentclass[12pt,preprint]{emulateapj}
\usepackage{natbib}

\shorttitle{}
\shortauthors{Mu\v{z}i\'c et al.}

\begin{document}
\bibliographystyle{apj}


\title{Discovery of Two Very Wide Binaries with Ultracool Companions\\ and a New Brown Dwarf at the L/T Transition}


\author{Koraljka Mu\v{z}i\'c\altaffilmark{1,*}, Jacqueline Radigan\altaffilmark{1}, Ray Jayawardhana\altaffilmark{1}, 
Valentin D. Ivanov\altaffilmark{2},
Jacqueline K. Faherty\altaffilmark{3,6},
Radostin G. Kurtev\altaffilmark{4},
Alejandro N\'u\~nez\altaffilmark{5,6},
Henri M. J. Boffin\altaffilmark{2},
Olivier Hainaut\altaffilmark{7},
Kelle Cruz\altaffilmark{5,6}, 
David Jones\altaffilmark{2},
Stanimir Metchev\altaffilmark{8},
Amy Tyndall\altaffilmark{2},
Jura Borissova\altaffilmark{4,9}}

\email{kmuzic@eso.org}

\altaffiltext{1}{Department of Astronomy \& Astrophysics, University of Toronto, 50 St. George Street, Toronto, ON M5S 3H4, Canada}
\altaffiltext{2}{European Southern Observatory, Alonso de C\'ordova 3107, Casilla 19, Santiago, 19001, Chile}
\altaffiltext{3}{Department of Astronomy, University of Chile, Camino El Observatorio 1515, Casilla 36-D, Santiago, Chile}
\altaffiltext{4}{Departamento de F\'isica y Astronom\'ia, Facultad de Ciencias, Universidad de Valpara\'iso, Av. Gran Breta\~na 1111, Playa Ancha, Casilla 53, Valpara\'iso, Chile}
\altaffiltext{5}{Department of Physics and Astronomy, Hunter College, City University of New York, 695 Park Ave, NY 10065, USA}
\altaffiltext{6}{Dept. of Astrophysics, American Museum of Natural History, New York, NY, 10024, USA}
\altaffiltext{7}{European Southern Observatory, Karl-Schwarzschild-Strasse 2, D-85748 Garching bei M\"unchen, Germany}
\altaffiltext{8}{Department of Physics and Astronomy, Stony Brook University, 100 Nicolls Rd, Stony Brook, NY 11794-3800}
\altaffiltext{9}{The Milky Way Millennium Nucleus, Av. Vicu\~{n}a Mackenna 4860, 782-0436 Macul, Santiago, Chile}
\altaffiltext{*}{Present address: ESO Chile \altaffilmark{2}}

\begin{abstract}
We present the discovery and spectroscopic follow-up of a nearby late-type L dwarf (2M0614+3950), and two 
extremely wide very-low-mass binary systems (2M0525-7425AB and 2M1348-1344AB), resulting from our search for common 
proper motion pairs containing ultracool components in the Two Micron All Sky Survey (2MASS) and the Wide-field 
Infrared Survey Explorer (WISE) catalogs. The near-infrared spectrum of 2M0614+3950 indicates a 
spectral type L$9 \pm 1$ object residing at a distance of  $26.0 \pm 1.8$  pc. The optical spectrum of 2M0525-7425A 
reveals an M$3.0 \pm 0.5$ dwarf primary, accompanied by a secondary previously classified as L2. The system has an 
angular separation of $\sim 44''$, equivalent to $\sim 2000\,$AU at distance of  $46.0 \pm 3.0$  pc.
Using optical and infrared spectra, respectively, we classify the components of 2M1348-1344AB as M$4.5 \pm 0.5$ 
and  T$5.5 \pm 1$.  The angular separation of $\sim 68''$ is equivalent to  $\sim 1400\,$AU 
at a distance of $20.7 \pm 1.4$ pc.  2M1348-1344AB is one of only six very wide (separation $>$ 1000 AU) systems containing late 
T dwarfs known to date. 
\end{abstract} 

\keywords{binaries:general - stars:low-mass, brown dwarf - stars: individual 
(2MASSJ06143818+3950357, 2M05254550-7425263, 2M05253876-7426008, 2M13480721-1344321, 2M13480290-1344071}

\section{Introduction}
\label{s1}
The formation and evolution of objects at the low-mass end of the initial
mass function is one of the most fundamental and yet not fully understood issues in our 
current picture of how stars form and evolve. 
The existence of brown dwarfs (BDs; $m<0.08\,$M$_{\odot}$) down to $\sim7\,$M$_J$
can be explained within the standard framework of star formation \citep{low76}. 
By including additional physics 
such as turbulence \citep{padoan&nordlund04}, dynamical interactions \citep{boffin98, bate09}, disk
fragmentation \citep{stamatellos09}, and/or photo-erosion from nearby bright stars
\citep{whitworth04}, one can create objects with even lower masses (down to $3\,$M$_J$).
The relative importance of these mechanisms in producing the Galactic substellar population is unclear. 
The key here is to confront predictions from the various scenarios with observational results.
One means of testing these theories is through the empirical characterization
of very-low-mass (VLM) multiple systems. As summarized in \citet{burgasser07}, 
these systems are observed to be less frequent, more tightly bound, and of higher mass ratios than
their more massive counterparts. 

Wide binaries  are of interest because their large separations (greater than 100 AU) 
and low binding energies provide direct constraints for formation models.
For example, wide binaries are expected to be disrupted during the ejection process. This
directly challenges the ejection model for the formation of VLM stars and BDs
\citep{reipurth&clarke01, bate&bonnell05}.
Several other models have been been successful in creating wide VLM binaries, e.g. by fragmentation
during the cluster formation \citep{bate09}, in fragmenting disks \citep{stamatellos09}, or by dynamical 
capture during star cluster dissolution phase \citep{kouwenhoven10, moeckel11}. The models by
\citet{bate09} and \citet{kouwenhoven10} predict the frequency of binaries with separations $>1000$AU to be
a few percent, compared with a fraction of $\sim10\%$ found observationally for nearby \citep{lepine07,longhitano10}
and young \citep{kraus09} stars. In star forming regions, the fraction of wide binaries 
appears to decrease with decreasing stellar mass \citep{kraus09}. While the number of known
wide binaries containing ultracool components is constantly growing as more searches are performed, 
they are mostly reported in an individual fashion, rather than in complete samples that would allow us
to asses their true frequency. 

Despite the effort that has been devoted to modeling over the past decade,
evolutionary tracks and atmosphere models for ultracool objects still suffer from uncertainties. 
Widely separated L and T dwarf companions to nearby stars are also valuable
for studying the evolution and atmospheres of brown dwarfs. The possibility for independent determination
of age, mass, and metallicity makes these brown dwarfs benchmark objects for calibration
of evolutionary models and atmospheric studies \citep{pinfield06, burgasser10, loutrel11}.

The transition between the spectral types L and T occurs over a
small temperature range of ∼200-300 K \citep{kirkpatrick00,dhan02,tinney03}, and is believed to be caused
by the depletion of condensate clouds, where the driving mechanism for the depletion is
inadequately explained by current cloud models \citep{ackerman01,burgasser02,knapp04}.
The complex dynamic behavior of condensate clouds of low temperature
atmospheres at the L/T transition is one of the leading problems in brown dwarf
astrophysics today \citep{tinney99, bailer-jones99,goldman05,artigau09,radigan12}. 

In this paper we report the discovery of a brown dwarf at the L/T transition, and two nearby wide
binaries consisting of M-dwarf primaries with early L-type and late T-type companions 
\footnote{One of the two binaries, 2M1348-1344, was
independently discovered by \citet{deacon12}}. 

\section{Search for wide binaries containing ultracool dwarfs}
\label{search}

\begin{figure*}
\centering
\includegraphics[width=16cm,angle=0]{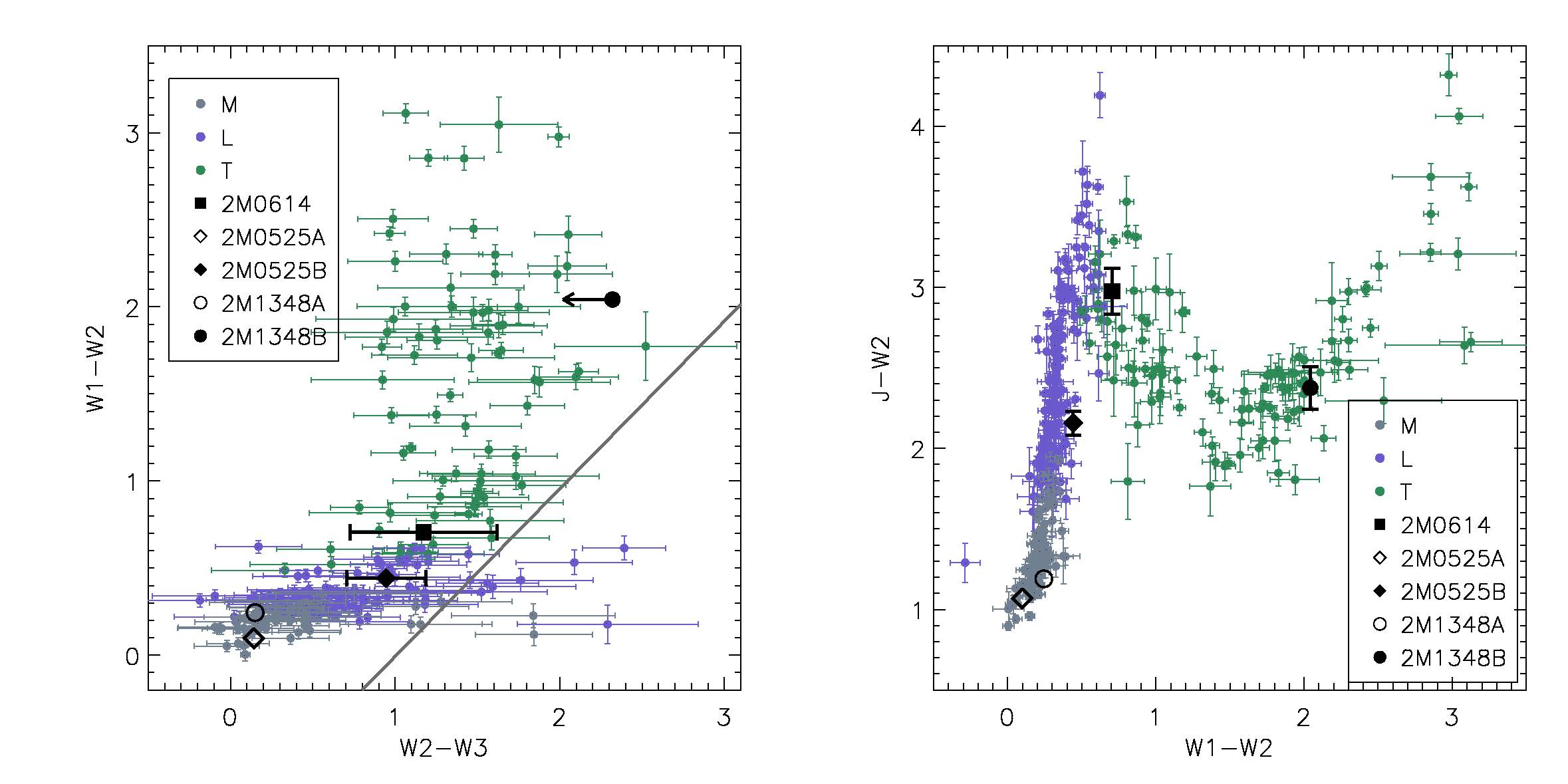}
\caption{2MASS and WISE color-color diagrams used for color selection of the MLT dwarf candidates. 
Diamonds and circles show the components of the two new wide binaries 2M0525-7425 and 2M1348-1344, respectively, 
while the filled square shows the new L9 dwarf 2M0614+3950.
Dots show the colors of the known MLT dwarfs from \citet{kirkpatrick11}. The solid line in the left panel is used to eliminate extragalactic sources to the right of the line from the bulk of the M, L, and T dwarfs to the left \citep{kirkpatrick11}. We do not plot error-bars in cases where they are comparable to or smaller than the plotting symbol.}
\label{ccds}
\end{figure*}

We performed a cross-match between the Two Micron All Sky Survey (2MASS) catalog and the 
Wide-field Infrared Survey Explorer (WISE) Preliminary Data Release catalog, in order 
to search for common proper motion pairs containing at least one very-low-mass component.
Following the successful approach 
of \citet{radigan08, radigan09},
the correlation of catalogs, calculation of proper
motions, and identification of co-moving stars was done in
overlapping sections of 4 deg$^2$ of the sky at a time. 
For every WISE source, we found the closest 2MASS match and computed 
proper motion vectors with uncertainties.
We then selected stars that had moved at $>3 \sigma$ level 
compared to all others within the search area. Stars within $120''$ of one another with proper 
motion amplitudes agreeing within $2\sigma$ and proper motion components agreeing within $1\sigma$ in either
right ascension or declination were flagged as potential binaries.
Towards the end of writing this paper, the new all-sky release of the WISE catalog became available. We have 
repeated the proper motion measurements for the 4 deg$^2$ fields around the three objects reported here.
The main reason for repeating these measurements is the improved astrometry for the faint sources 
in the new catalog. The uncertainties in declination in the WISE preliminary catalog
were computed by adding $0.5''$ in quadrature to the extraction measurement uncertainty to reflect the impact of the declination bias error known to affect a large fraction of sources fainter than $W1\sim\,$13$\,$mag. This has been corrected in the 
new release. Thus, while the main search for binaries has been performed using the preliminary version,
listed proper motions and photometry are based on the all-sky release of the WISE catalog.   

\citet{kirkpatrick11} recently published a compilation of 2MASS and WISE photometry for $\sim350$ early 
M- to late T-dwarfs. Mid-infrared WISE photometry is particularly suitable for L/T dwarf selection, because 
of their red $W1-W2$ colors, which are almost unique among astrophysical sources. Additionally, 
$W2-W3$ color can be used to distinguish between BDs on the one side, and AGB stars and dust-obscured 
galaxies on the other \citep{eisenhardt10}. 
In the following, we describe the procedure for selection of the candidate binary systems suitable
for spectroscopic follow-up.
First, we required A, B or C photometric quality flags for 
$J$, $H$, $K$, $W1$, $W2$ and $W3$ photometry for all the components of the 
potential binary systems. After this initial flagging, we were left with $\sim 11000$ objects. 
We then rejected all candidates for which at least one of the components does not satisfy the
following conditions: located to the left of the solid line in the $W2-W3$ vs. $W1-W2$ diagram (Figure~\ref{ccds} left),
$J-H$ between $-1$ and $2$, $W2-W3>-0.5$, $J-W2>0.5$, and $W1-W2>0.35$. The latter condition was applied
to discard a large group of contaminants that occupy the color space with $W1-W2\lesssim0.3$ and $J-W2<6$ in the right panel
of Figure~\ref{ccds}. Relaxing this condition to $W1-W2>0$ results in $\sim1200$ candidates. By applying the
cut at $W1-W2=0.35$, we were left with only 19 candidates. Clearly, this restriction makes our search
insensitive to essentially all M-M pairs, and a large portion of L-type companions earlier than L6 will be
discarded as well.
After applying the color cuts on either of the components 
of a candidate binary system, we inspected the remaining 19 candidates visually, in order 
to remove false positives. In most cases these contain blends of unresolved source pairs in WISE, due
to significantly lower resolution when compared with 2MASS. 

Two color-color diagrams used for the selection are shown in Figure~\ref{ccds}. Color-coded dots show 
the colors of M, L and T dwarfs taken from \citet{kirkpatrick11}, while the black symbols show the objects 
reported in this paper. In the following subsections we describe the three discoveries reported in this paper, 
along with the known parameters obtained from the literature.

2MASS and WISE photometry for the objects reported in this paper are listed in Table~\ref{T_phot}. Table~\ref{T_prop} 
contains the proper motions and other physical properties of the objects discussed in the next sections. 

\subsection{2M0614+3950}
2M06143818+3950357 (hereafter 2M0614+3950) was at first identified as a companion to 2M06143791+3951202. 
With $W2-W3~\approx 3$, $W1-W2~\approx 0$, and $J-W2~\approx 0.75$, the latter object is marginally 
consistent with being an early M-dwarf. However, an M0 dwarf as faint as $J=16.3$ would have to be 
at a distance larger than 1kpc. By checking the optical catalogs of the region, we found that the optical 
source matching 2M06143791+3951202 has essentially the same position as the 2MASS source 
(offset of $\sim 0.2''$ from GSC2.2 and USNO-B1 catalog positions), whereas at the measured proper 
motion we would expect an offset of $\sim2.5''$ between the GSC2.2 and 2MASS, and $>5''$ between 
USNO-B1 and 2MASS. Furthermore, in the optical images the position reported by WISE appears to be 
located between the source matching the 2MASS position and another faint (R$\sim19$) source not 
detected by 2MASS. We thus conclude that the WISE source identified here as a potential
primary of a binary system is most probably a blend.
The suspected secondary, however, has colors that are consistent with a late-L or an early-T ultracool
dwarf, and was therefore included in our spectroscopic follow-up.

\subsection{2M0525-7425}
A common proper motion pair here denoted as 2M0525-7425 consists of the primary 2M05254550-7425263 and the 
secondary 2M05253876-7426008. Both objects were reported by \citet{kirkpatrick10} as part of a large-area
proper motion survey. The secondary was observed with GMOS/Gemini during their spectroscopic follow-up, and was 
assigned the optical spectral type L2. The primary was also reported by \citet{subasavage05} as a high-proper motion 
source at a photometric distance of 28.7~pc. Despite the matching proper motions, the pair was never reported as 
a wide binary candidate; the primary lacked spectral classification.

\subsection{2M1348-1344}
A common proper motion pair here denoted as 2M1348-1344 consists of the primary 2M13480721-1344321 
(also known as LP 738-14, NLTT 35266, LHS 2803) and the secondary 2M13480290-1344071. 
\citet{reid03} published photometry of the primary, from which they estimate distance of 20.9~pc. 
\citet{casagrande08}, again based on photometry, estimated an effective temperature of $T_{\mathrm{eff}} = 2942 \pm 58\,$K. 
The secondary has colors consistent with late T-dwarfs (Figure~\ref{ccds}).

\begin{deluxetable*}{lccccccccc}
\tabletypesize{\tiny}
\tablecaption{2MASS and WISE Photometry}
\tablewidth{0pt}
\tablehead{\colhead{name} & 
	   \colhead{$\alpha$\tablenotemark{a}} & 
	   \colhead{$\delta$} & 
	   \colhead{J} & 
	   \colhead{H} & 
	   \colhead{K} &
	   \colhead{W1} & 
	   \colhead{W2} & 
	   \colhead{W3} & 
	   \colhead{W4}}
\tablecolumns{10}
\startdata
2M1348-1344A & 13 48 07.22 & -13 44 32.1 & $10.41 \pm 0.02$ & $9.94 \pm 0.02$ & $9.66 \pm 0.02$ &
				       $9.46 \pm 0.02$ & $9.24 \pm 0.02$ & $9.11 \pm 0.03$ & $8.72 \pm 0.33$\\
2M1348-1344B & 13 48 02.91 & -13 44 07.2 & $16.48 \pm 0.12$ & $16.09 \pm 0.17$ & $>16.45$ &
				      $16.15 \pm 0.07$ & $14.18 \pm 0.05$ & $>12.14$ & $>9.14$\\
2M0525-7425A & 05 25 45.50 & -74 25 26.3 & $10.03 \pm 0.02$ & $9.42 \pm 0.02$ & $9.21 \pm 0.02$ & 
				       $9.08 \pm 0.02$ & $8.97 \pm 0.02$ & $8.82 \pm 0.02$ & $8.53 \pm 0.15$\\
2M0525-7425B & 05 25 38.76 & -74 26 00.8 & $15.71 \pm 0.07$ & $14.97 \pm 0.10$ & $14.43 \pm 0.10$ & 
				      $14.02 \pm 0.03$ & $13.57 \pm 0.03$ & $11.97 \pm 0.12$ & $>9.46$\\
2M0614+3950  & 06 14 38.18 & +39 50 35.7 & $16.59 \pm 0.14$ & $15.60 \pm 0.12$ & $15.02 \pm 0.12$ & 
				      $14.33 \pm 0.03$ & $13.66 \pm 0.04$ & $12.19 \pm 0.34$ & $>9.22$
\enddata
\label{T_phot}
\tablenotetext{a}{coordinates from 2MASS}
\end{deluxetable*}

\section{Spectroscopic follow-up}
\label{followup}

\subsection{EFOSC2/NTT}
We obtained optical spectroscopy of 2M0525-7425A (December 23 2011; program ID 184.C-1143(E)) and 
2M1348-1344A (February 29 2012; 088.D-0573(A)) with the ESO Faint Object Spectrograph and Camera (EFOSC2; \citealt{buzzoni84}) on the ESO New Technology Telescope (NTT).
Observations were taken in long-slit mode with a $1''$ slit and grism \#05, covering wavelengths 5200\AA-9350\AA\
with a resolution of 16.6\AA. The data were flat-fielded, extracted, wavelength- and flux-calibrated using the EFOSC2 pipeline provided by ESO.
The EFOSC2 grism \#05 used in this work introduces stong fringing redwards of about $\sim$7200\AA. In the case of 2M1348-1344A the afternoon dome flats adequatly removed the fringing, while the effect remains present in the spectrum of 2M0525-7425A. The imperfect flat-fielding is due to telescope/instrument flexure, which may cause a sligt positional mismatch between the fringes in the dome flats and science images. The signal-to-noise ratio, as measured by the iraf task {\it splot} varies between 5 and 18 for 2M0525-7425A, and 5 and 23 for 2M1348-1344A.  


\subsection{SpeX/IRTF}
We obtained near-infrared spectroscopy of 2M0614+3950 on December 8 2011 with SpeX \citep{rayner03} on the NASA Infrared Telescope Facility (IRTF) on Mauna Kea. We used the low-resolution prism mode of SpeX to provide a wavelength coverage of 0.8-2.5 microns in a single order and an average resolution of R$\sim120$ with a 0$\farcs$5-wide slit.
The spectrum was acquired using the 0.5-wide slit aligned at parallactic angle, with an airmass that ranged from 1.08 to 1.17 during the observations. The target was observed with multiple exposures, dithering in an ABBA pattern along the slit, with a total integration time of 2880 sec. After the target observation, a nearby A0 star at a similar airmass to the target was observed for both telluric and flux calibration. Internal flat field and argon arc lamp calibration frames were obtained at the target position for pixel response and wavelength calibration. The spectral data was extracted, flat-fielded, wavelength calibrated, telluric corrected, and flux-calibrated with the {\it Spextool} reduction package \citep{cushing04,vacca03}.
The signal-to-noise ratio of the final spectrum varies between 6 and 36.

\subsection{FIRE/Magellan}
We obtained near-infrared spectroscopy of 2M1348-1344B on March 20 2012 with the Folded-port InfraRed Echelette (FIRE;
\citealt{simcoe08, simcoe10}) spectrograph on the 6.5-m Magellan Baade Telescope. 
In its high throughput prism mode, FIRE covers the wavelength range from 0.8 to 2.5 $\mu$m at a resolution
ranging from R=500 at $J$-band to R=300 at $K$-band for a slit width of $0.6''$. A series of ABBA nod pairs taken with exposure times of 120s per position were used giving a total exposure time of 480s. The spectrograph
detector was read-out using the 4-amplifier mode at “high gain” (1.2 counts per e$^-$) with
the SUTR sampling mode. We also obtained exposures of a variable voltage quartz lamp
for flat-fielding purposes and neon/argon arc lamps were used for
wavelength calibration. The A0 dwarf star HIP70419 was used for telluric and flux calibration.
Data were reduced with the new FIRE pipeline tools implemented in IDL and written by
R. Simcoe, J. Bochanski and M. Matejek. The signal-to-noise ratio of the final spectrum varies between 4 and 14.

\section{Results and discussion}
\label{results}

\subsection{2M0614+3950: New brown dwarf at L/T transition}
\label{Sec_2M0614}

\begin{figure}
\centering
\includegraphics[width=8.5cm,angle=0]{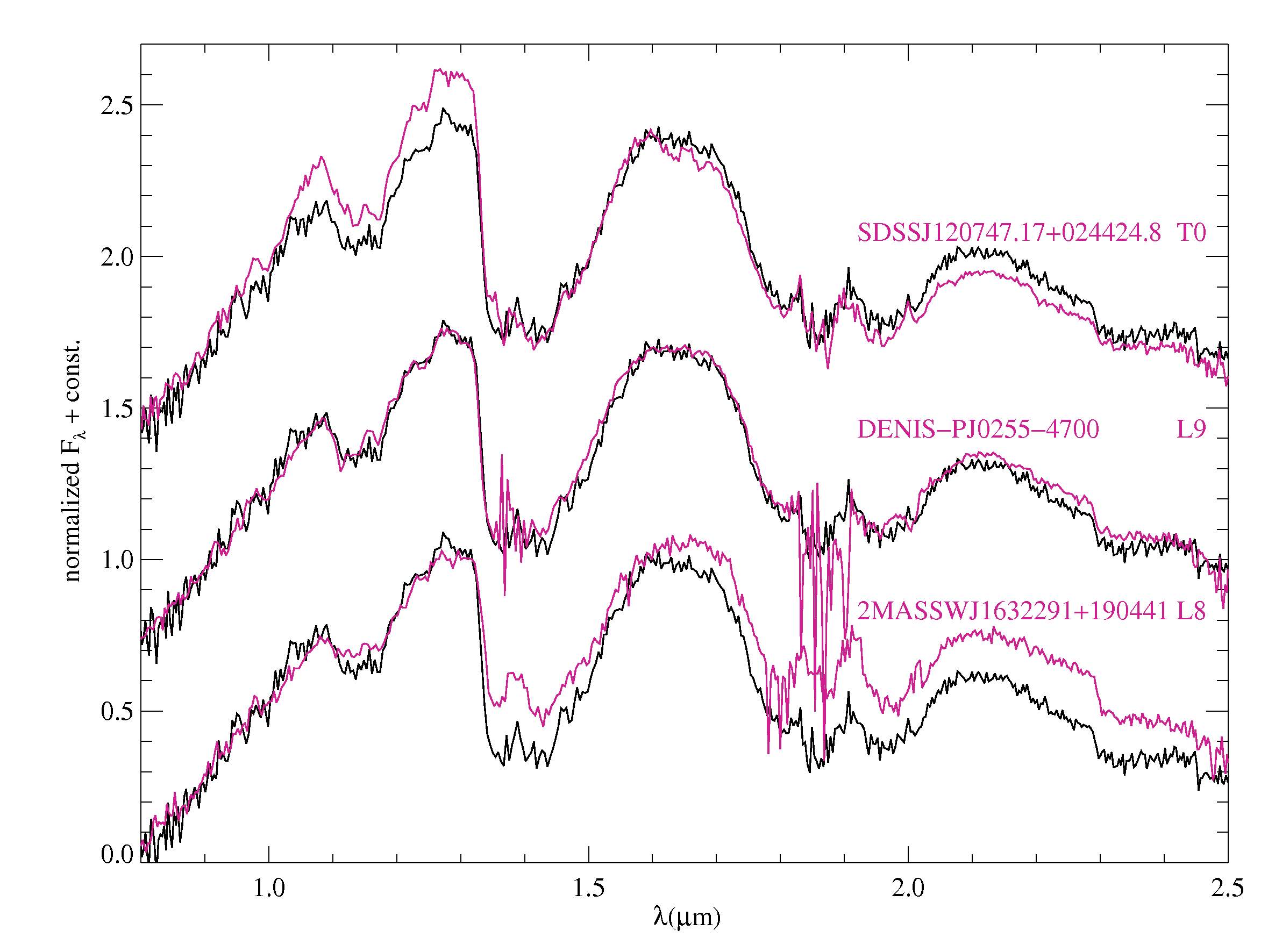}
\caption{SpeX spectrum of 2M0614+3950 shown in black, with the standard spectral sequence of ultracool
dwarfs at the L/T transition \citep{burgasser06, burgasser07, looper07}. All spectra were normalized at $1.6 \mu$m.}
\label{2M0614}
\end{figure}

\begin{figure}
\centering
\includegraphics[width=9cm,angle=0]{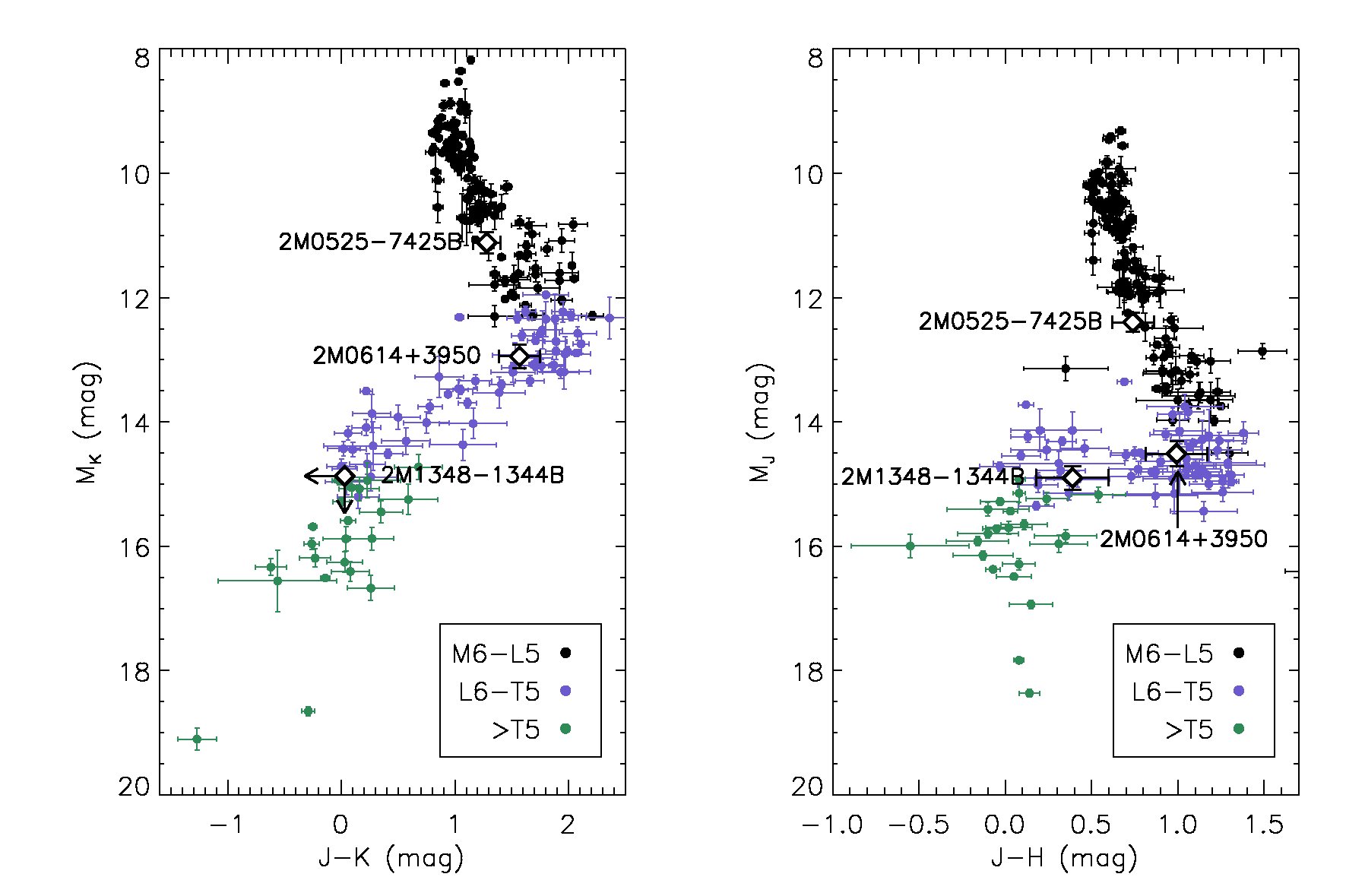}
\caption{Color-magnitude diagrams showing ultracool dwarfs with parallaxes from Table~10 of \citet{dupuy12} with
errors $\leq0.5\,$mag. Objects at the L/T transition are shown as purple circles. The absolute magnitude of 2M0614+3950 (L9) was calculated assuming a distance of d=$26.0 \pm 1.8\,$pc; for 2M1348-1344B (T5.5) we use d=$20.7 \pm 1.4\,$pc,  and for 2M0525-7425 (L2) d=$46.0 \pm 3.0\,$pc. }
\label{cmdNIR}
\end{figure}

The spectrum of 2M0614+3950 is well matched by the L9 near-infrared spectral template. In Figure~\ref{2M0614} we show the comparison
of the spectrum of our target (black) and the SpeX NIR standards from L8 to T0 \citep{burgasser06, burgasser07, looper07}.

\citet{dupuy12} provide a comprehensive update to the absolute magnitudes of ultracool dwarfs with measured parallaxes as a function of spectral type, for a wide variety of broad-band filters including 2MASS and WISE passbands. We use the relations between absolute J, H, K, W1, and W2 magnitudes and spectral type to calculate the distance to 2M0614+3950. The distance is calculated as a weighted average of the five estimates, with the weights equal to the inverse square of the individual uncertainties. These include the rms uncertainty of the polynomial fits 
given in \citet{dupuy12}, uncertainty of one spectral subtype, and the photometric errors.
For an object of spectral type L9 we obtain { d=$26.0 \pm 1.8\,$pc}. 
However, based on the recent analysis of the parallaxes for 70 nearby ultracool dwarfs, 
\citet{faherty12} point out that the brightening across the L/T transition is more pronounced in reality
than when represented by the polynomial fits
of absolute magnitudes versus spectral type. Therefore, in addition to the polynomial fits, they
calculate linear piecewise functions for three spectral type regions. The distance obtained from the
polynomial fits in \citet{faherty12} is in agreement with the above estimate. The piecewise fits, however, 
locate the object somewhat closer, at $\sim22\,$pc (average from JHK estimates). We note that
distances for objects at the L/T transition are uncertain as we are still investigating their significant scatter in luminosity.  

Figure~\ref{cmdNIR} shows the position of 2M0614+3950 in NIR color-magnitude diagrams, alongside ultracool dwarfs with
parallaxes and errors $\leq 0.15$mag from \citet{dupuy12}. 
2M0614+3950 is located in the region mainly populated by ultracool dwarfs of spectral type between $\sim$L7 and $\sim$T2. Since the distance
to the object was calculated assuming the absolute magnitude of an L9 dwarf, its placement in the correct magnitude range is assured. 
The colors of 2M0614+3950 are consistent with the L7 - T2 spectral range and do not show any peculiarities.


Using the relations between effective temperature and spectral type \citep{stephens09}, and absolute bolometric
magnitude and spectral type \citep{burgasser07}, we obtain 
 $T_{\mathrm{eff}} \approx 1350 \pm 110\,$K, and log(L/L$_{\odot}$)=$-4.61 \pm 0.10$.
The uncertainties are derived by adding in quadrature the rms error
of the polynomial fits and the error resulting from the uncertainty of one spectral subtype.
Compared to the DUSTY theoretical
isochrones \citep{chabrier00,baraffe02}, this would correspond to an object with a mass of 
$\sim0.04 - 0.07\,M_{\odot}$, for the ages of 1-10 Gyr.
A summary of the observational and physical properties of 2M0614+3950 is given in Table~\ref{T_prop}.

\subsection{2M0525-7425: New wide M/L binary}
\label{2M0525_results}

\begin{figure}
\centering
\includegraphics[width=7.0cm,angle=0]{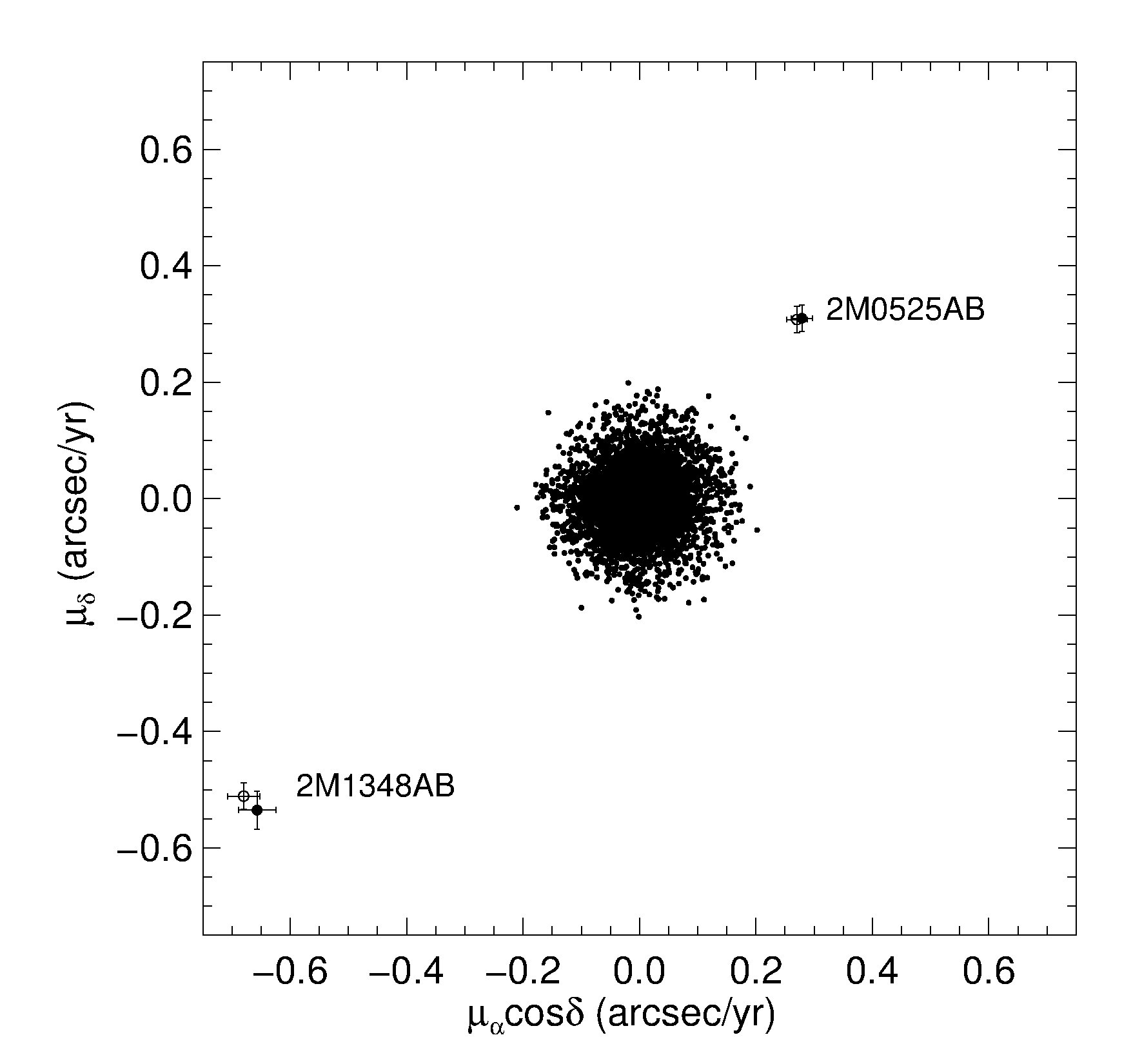}
\caption{Proper motions of the sources within 1 deg$^2$ around 2M0525-7425 and 2M1348-1344. The components of the wide binary candidates 
2M0525-7425 and 2M1348-1344 are marked with the open (primary) and filled (secondary) circles with 
corresponding errorbars.}
\label{pm}
\end{figure}

\begin{figure}
\centering
\includegraphics[width=8.5cm,angle=0]{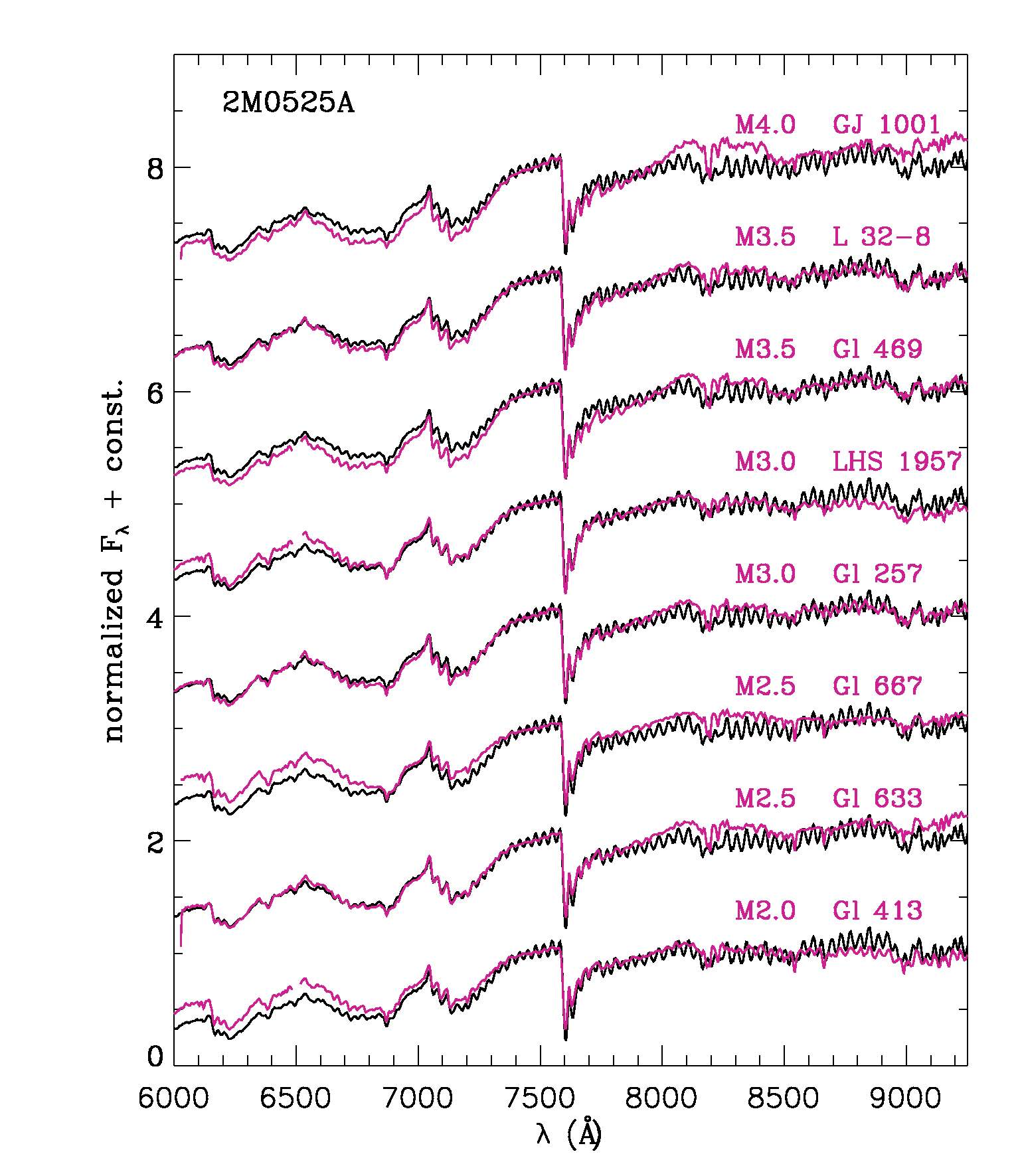}
\caption{EFOSC2 spectrum of 2M0525-7425A shown in black, with the spectral sequence of early
M-dwarfs from {\tt dwarfarchives.org} overplotted in purple. The comparison spectra were chosen to 
have broad wavelength coverage similar to our data, and were smoothed to match the resolution of the object spectrum. 
All spectra were normalized at 7500 \AA.}
\label{2M0525A}
\end{figure}

Figure~\ref{pm} shows proper motions of the sources within 1 deg$^2$ around 2M0525-7425 and 2M1348-1344. The matching radius 
was set to $2''$ in order to discard mismatches caused by faint 2MASS sources not detected in WISE. 
The components of the wide binary 2M0525-7425 are marked 
with open (primary) and filled (secondary) circles with errorbars, which are calculated
by combining the astrometric uncertainties of the sources in 2MASS and WISE catalogs.

In Figure~\ref{2M0525A} we show the comparison of the spectrum of the primary (black) and CTIO-4m 
and CTIO-1.5m spectra of dwarfs with spectral types from M2 to M4 (purple). The comparison spectra 
were taken from {\tt www.dwarfarchives.org}, and were chosen to match the EFOSC2 wavelength coverage. 
The spectrum of 2M0525-7425A is well matched by the spectra of M2.5 - M3.5 dwarfs. We therefore assign 
it a spectral type M$3 \pm 0.5$. 

\citet{reid&cruz02} computed color-magnitude relations in the ($M_V$, $V-K$), ($M_V$, $V-I$), and ($M_I$, $I-J$) 
planes for nearby stars with well-determined trigonometric parallaxes, which allows us to determine a 
photometric parallax for 2M0525-7425A. We use JK photometry from 2MASS, V=13.57 (NOMAD catalog; \citealt{zacharias05}), 
and I=11.3 (DENIS catalog). Combining the results from the three relations in \citet{reid&cruz02}, 
we obtain a distance of d=$43.9 \pm 4.3\,$pc. 
The distance is calculated as a weighted average of the three estimates, with the weights equal to the inverse square of the 
individual uncertainties resulting from the scatter of the color-magnitude relations given in \citet{reid&cruz02}. 
The photometric uncertainties are either negligible compared to the fit errors ($IJHK$), or not available ($V$), and therefore were not 
included in the calculation.

The distance to 2M0525-7425B can be estimated in the same way as for 2M0614+3950 (see Section~\ref{Sec_2M0614}). 
The secondary was classified as an L2 dwarf \citep{kirkpatrick10}: we estimate that it lies at a distance 
{ d=$48.0 \pm 4.3\,$pc}. The excellent agreement between independent distance estimates for the two objects 
gives support to the wide-binary nature of the pair. In the following, we adopt the 
weighted average  d=$46.0 \pm 3.0$$\,$pc  as the distance to the 2M0525-7425 binary, 
with the weights equal to the inverse square of the component distance uncertainties.
At this distance, the projected separation of $43.9'' \pm 0.1''$ is 
equivalent to  $2020 \pm 130\,$AU. 
The uncertainty of the projected separation combines the astrometric errors of the binary components from 2MASS.

Using the relations between effective temperature and spectral type \citep{stephens09}, and absolute bolometric
magnitude and spectral type \citep{burgasser07}, we obtain  $T_{\mathrm{eff}} \approx 1970 \pm 120\,$K, 
and log(L/L$_{\odot}$)=$-3.79 \pm 0.11$  for the L2 companion. 
The spectral type for this source is adopted from \citet{kirkpatrick10},
which gives no formal errors on spectral types. We therefore assume an uncertainty of 0.5 subtypes, that is equal to the 
step in their spectral typing scheme. The resulting uncertainty was added in quadrature to the scatter of the polynomial fits.
For the primary, we use relations between effective 
temperature and color determined by \citet{casagrande08} for a large sample of nearby M dwarfs with 
accurate optical and infrared photometry. We calculate $T_{\mathrm{eff}}$ as the average of the 
estimates based on six optical-to-near-infrared colors (V and R versus J, H, and K), with the
uncertainty equal to the standard deviation of these six estimates. The R magnitude 
originates from the HST Guide Star Catalog-II (GSC-II)~\footnote{Photographic R$_F$ magnitude from GSC-II 
equivalent to the Johnson-Cousins R$_C$ \citep{reid91, kirk09}.}. We obtain   $T_{\mathrm{eff}} 
\approx 3320 \pm 50\,$K.  From the relation between the 2MASS and bolometric magnitudes we obtain
log(L/L$_{\odot}$)=$-1.60 \pm 0.03$. The uncertainty takes into account the scatter of the linear
relations between $JHK$ and bolometric magnitudes \citep{casagrande08}, photometric errors, and the distance uncertainty.

\citet{lepine07b} defined the metallicity dependent index $\zeta$ using the CaH2, CaH3 and TiO5 molecular
band heads in the optical, recently re-calibrated by \citet{dhital12}. From the EFOSC2 spectrum of 2M0525-7425A
we find $\zeta_{L}$=0.94 and $\zeta_{D}$=0.89, 
placing the object among the dwarf metallicity class. 
However, while the $\zeta$-index can be very useful for discriminating between different metallicity
classes, its correlation with [Fe/H] within the dwarf metallicity class ($\zeta > 0.825$) is weak and
shows a significant scatter \citep{lepine12}.
The location of M dwarfs in the ($V - K_S$) -- $M_{Ks}$ color-magnitude diagram has been shown to correlate with metallicity
\citep{bonfils05,johnson09,schlaufman10}. Using the relation from \citet{schlaufman10}, who improved upon previous 
calibrations, and adopting the average distance to the binary
 d=$46.0 \pm 3.0$$\,$pc, we obtain [Fe/H]=$-0.01 \pm 0.03$ for 2M0525-7425A. 
The final uncertainty combines the uncertainties of the distance and 
the $K_S$ magnitude, while the error of the $V$ magnitude is not available.
Note that \citet{neves12} find the rms dispersion of the \citet{schlaufman10} relation 
to be $0.19 \pm 0.03\,$dex.

\citet{west08} showed that the decrease in
the fraction of magnetically active stars (as traced by H$\alpha$ emission) as a function of 
the vertical distance from the Galactic plane can be explained by thin-disk dynamical heating and a
rapid decrease in magnetic activity. The timescale for this rapid activity decrease changes according 
to the spectral type.  
From the lack of H$\alpha$ emission in the spectrum of 2M0525-7425A, we place a lower limit for its age at 1.5 Gyr. 

Assuming an age of 1-10 Gyr, the component masses can be estimated from evolutionary models. 
By comparing the derived absolute JH$K_S$ magnitudes with the BCAH98 evolutionary models \citep{baraffe98}, we
find a mass of $0.46 \pm 0.01\,$M$_{\odot}$ for the primary. The adopted uncertainty was determined by 
varying the absolute magnitudes within their corresponding uncertainties.
This result is in agreement with the value $0.45  \pm 0.01\,$M$_{\odot}$ obtained using the empirical 
mass-luminosity relationship of \citet{delfosse00}.
For the secondary, we find a mass of $0.070^{+0.005}_{-0.010}\,$M$_{\odot}$, with an error reflecting the model grid spacing. 
The mass ratio of 2M0525-7425AB is $0.15 \pm 0.02$, which is relatively low for a low-mass pair. The mass
ratios are discussed in Section~\ref{q}.
A summary of observational and physical properties of 
2M0525-7425AB is given in Table~\ref{T_prop}.


\subsection{2M1348-1344: New wide M/T binary}
\label{Sec_2M1348}

\begin{figure*}
\centering
\plottwo{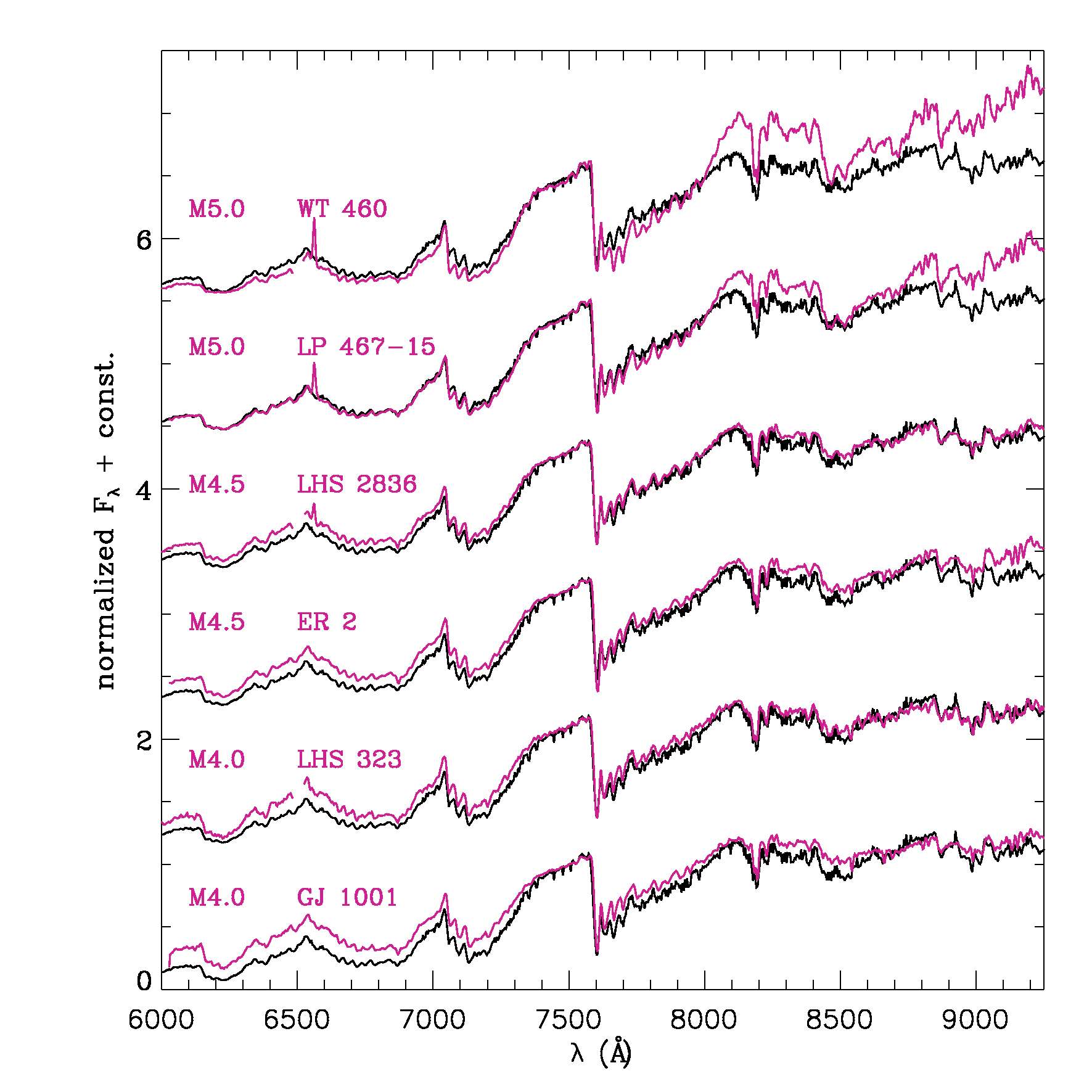}{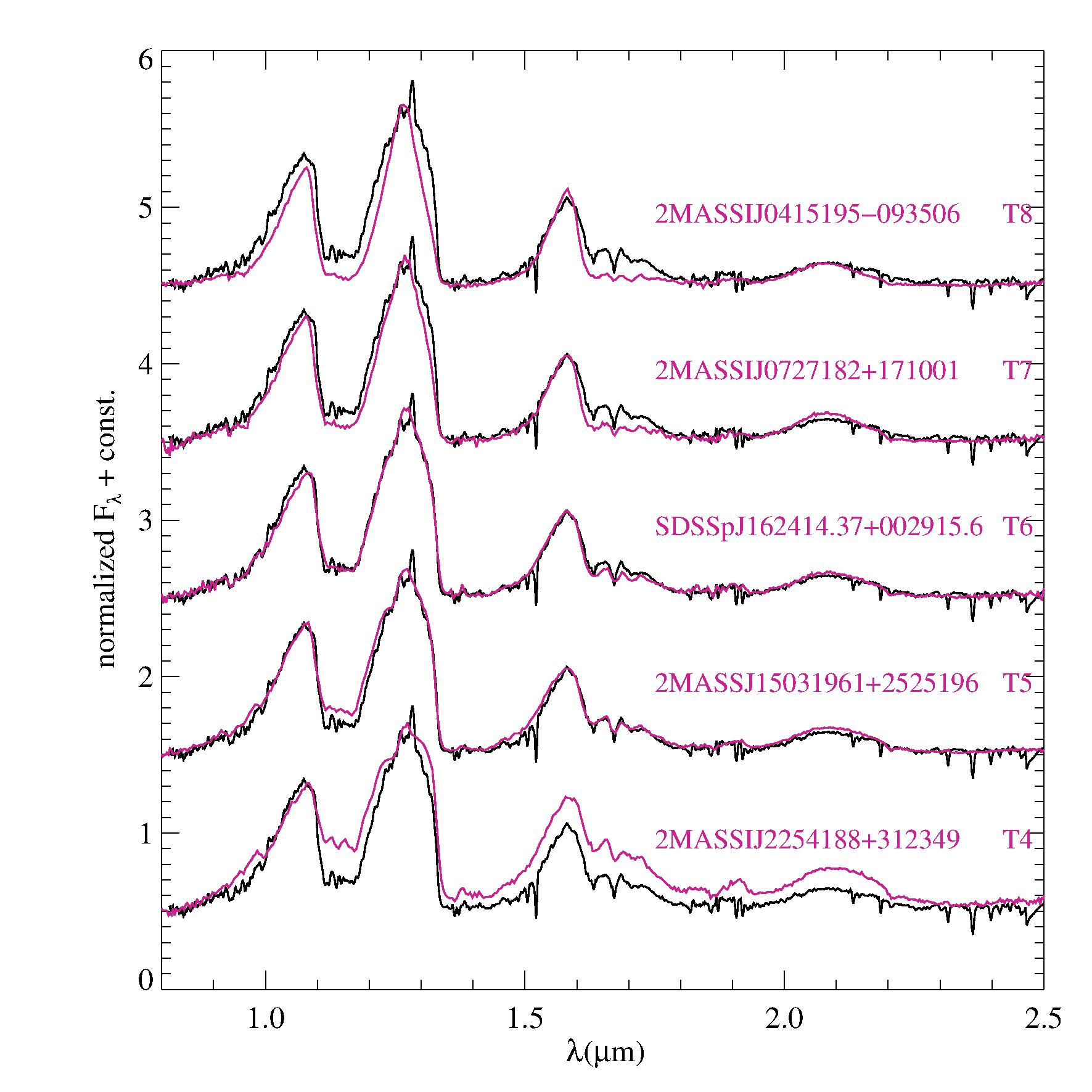}
\caption{
{\bf Left:} EFOSC2 spectrum of 2M1348-1344A shown in black, with the M-dwarf spectral sequence
from dwarfarchives.org overplotted in purple.
The comparison spectra were chosen to have broad wavelength coverage similar to our data, and were smoothed to match the resolution of the object spectrum. All spectra were normalized at 7500 \AA.
{\bf Right:} FIRE spectrum of 2M1348-1344B shown in black, with the standard spectral sequence of T dwarfs 
\citep{burgasser04, burgasser06}. The higher-resolution FIRE spectrum was smoothed to match the resolution of the spectral standards.
All spectra were normalized at $1.25 \mu$m.}
\label{2M1348_fig}
\end{figure*}

In the left panel of Figure~\ref{2M1348_fig} we show the comparison
of the EFOSC2 spectrum of the primary (black) and the CTIO-4m and CTIO-1.5m spectra of dwarfs with 
spectral types from M4 to M5 (purple). The comparison spectra were taken from www.dwarfarchives.org, 
and were chosen to match the EFOSC2 wavelength coverage. Based on this plot, we assign 2M1348-1344A 
a spectral type of M$4.5 \pm 0.5$. 
The FIRE spectrum of the secondary is shown in black in the right panel of Figure~\ref{2M1348_fig}, 
along with the SpeX NIR standards from T4 to T8 \citep{burgasser04, burgasser06}. 
Spectral templates T5 and T6 both provide good fit to the spectrum of 2M1348-1344B. 
As a check, we calculate the spectral
indices from \citet{burgasser06}, that are based on the strength of the major H$_2$O and CH$_4$ bands in the near-infrared. 
Five indices yield spectral types from T5 to T7, with the average of T5.7. 
We type 2M1348-1344B as T$5.5 \pm 1.0$.

The distance to the primary can be calculated in the same way as for 2M0525-7425A (Section~\ref{2M0525_results}), 
adopting V=15.1 and I=12.1 
\citep{reid03}. This gives us d$\,=\,21.5 \pm 2.1\,$pc. The distance to the T-dwarf was calculated 
in the same way as for the L2 and L9 dwarfs from the previous two sections, based
on J, H, W1 and W2 magnitudes and by adopting the spectral type T5.5. We obtain  d$\,=\,20.0 \pm 2.0\,$pc .
The excellent agreement between the independent distance estimates for the two objects supports
a binary classification. In the following we adopt the weighted average  d=$20.7 \pm 1.4$$\,$pc  
as the distance to 2M1348-1344 binary. 
At this distance, the projected separation of $67.6'' \pm 0.3''$ is equivalent to   $1400 \pm 90\,$AU.  

Following the same procedure as in Section~\ref{2M0525_results}, we obtain 
$T_{\mathrm{eff}} \approx 2960 \pm 20\,$K, and log(L/L$_{\odot}$)=$-2.42 \pm 0.03$ for the primary.
The activity lifetime for M4 dwarfs is $4.5 \pm 0.5$ Gyr. From the lack of H$\alpha$ emission 
in the spectrum of 2M1348-1344A, we place a lower limit for its age at 4 Gyr. 
The $\zeta$-index of 2M1348-1344A is found to be very close to unity 
($\zeta_{LRS}=1.08$, $\zeta_{D}=1.06$), placing the object into the dwarf metallicity class.
The photometric relation of \citet{schlaufman10} results in [Fe/H]=$-0.33 \pm 0.08$.
Here we have to take into account the rms dispersion of the relation that equals
$0.19\pm 0.03\,$, as calculated
by \citet{neves12}.
Our result for the metallicity of 2M1348-1344A is therefore in agreement with the 
values found for other M-dwarfs in the solar neighbourhood, 
which have the mean [Fe/H]=$-0.17 \pm 0.07$ \citep{schlaufman10}.
The estimate for the mass of the primary gives $0.18 \pm 0.01\,$M$_{\odot}$ (BCAH98, age 5-10 Gyr). 
The same number is obtained by following the relationship of \citet{delfosse00}. 
For the T5.5 secondary, we obtain  $T_{\mathrm{eff}} \approx 1070 \pm 130\,$K, and 
log(L/L$_{\odot}$)=$-5.03 \pm 0.21$. Using COND03 \citep{baraffe03} evolutionary models, we find the mass of 
the secondary to be $0.04 \pm 0.01\,$M$_{\odot}$ for ages 5-10 Gyr. As in the case of 2M0525-7425, 
we find a relatively low mass ratio of $0.22 \pm 0.06$. 
A summary of the observational and physical properties of 2M1348-1344
is given in Table~\ref{T_prop}.

2M1348-1344B is one of only a handful of known late T ($\geq$T5) wide companions to low mass stars, and one
of only six late T dwarfs separated by more than 1000 AU from their primary stars. The other objects include 
Gl 570D (K4/M1.5/M3/T7 system, K4 and T7 separated by 3600 AU; \citealt{burgasser00}), 
$\epsilon$~Indi Bb (K5/T1/T6 system, K5 separated by 1460 AU from the T1-T6 close binary; \citealt{scholz03, mccaughrean04}), 
Ross 458C (M0/M7/T8.5, M0 and T8.5 separated 1170 AU; \citealt{scholz10, goldman10, burgasser10b, burningham11}), 
Hip 73786B (K5/T6p with separation of 1200 AU; \citealt{scholz10, murray11}), and
G204-39B (M3/T6.5 separated by 2685 AU; \citealt{faherty10}).

\begin{deluxetable*}{lccccc}
\tabletypesize{\scriptsize}
\tablecaption{Physical properties of 2M1348-1344AB, 2M0525-7425AB, and 2M0614+3950}
\tablewidth{0pt}
\tablehead{\colhead{} &
	   \colhead{2M1348-1344A} & 
	   \colhead{2M1348-1344B} &
	   \colhead{2M0525-7425A} & 
	   \colhead{2M0525-7425B} &
	   \colhead{2M0614+3950}}
\tablecolumns{6}
\startdata
 2MASS designation & 13480721-1344321 & 13480290-1344071 & 05254550-7425263 & 05253876-7426008 & 06143818+3950357\\
 $\mu_{\alpha} cos\delta$ (mas yr$^{-1}$) & $-680 \pm 28$ & $-657 \pm 23$ & $270 \pm 18$ & $279 \pm 18$ & $-46 \pm 21$\\
 $\mu_{\delta}$ (mas yr$^{-1}$) 	  & $-511 \pm 32$ & $-535 \pm 33$ & $308 \pm 23$ & $310 \pm 23$ & $-265 \pm 23$\\
 Spectral type & M$4.5 \pm 0.5$\tablenotemark{b} & T$5.5 \pm 1$\tablenotemark{c}  & M$3.0 \pm 0.5$\tablenotemark{b} & 
                 L2\tablenotemark{a}\tablenotemark{b} & L$9 \pm 1$\tablenotemark{c} \\
 Distance (pc) & $21.5 \pm 2.1$ & $20.0 \pm 2.0$ & $43.9 \pm 4.3$ & $48.0 \pm 4.3$ & $26.0 \pm 1.8$\\
 $T_{\mathrm{eff}}$(K) & $2960 \pm 20$ & $1070 \pm 130$ & $3320 \pm 50$ & $1970 \pm 120$ & $1350 \pm 110$ \\
 Mass (M$_{\odot}$) & $0.18 \pm 0.02$ & $0.04 \pm 0.01$ & $0.46 \pm 0.01$ & $0.070^{+0.005}_{-0.010}$ & $0.04-0.07$\\ 
 Age (Gyr) & $\geq 4.0$    & & $\geq 1.5$  & & \\
 $[$Fe/H$]$\tablenotemark{d} & $-0.33 \pm 0.08$ &  & $-0.01 \pm 0.03$ & &\\
 Angular separation ($''$)& \multicolumn{2}{c}{$67.6 \pm 0.3$}  & \multicolumn{2}{c}{$43.9 \pm 0.1$} &\\
 Projected separation (AU) & \multicolumn{2}{c}{$1400 \pm 90$} & \multicolumn{2}{c}{$2020 \pm 130$} &\\
 Binding energy (10$^{41}$ erg) & \multicolumn{2}{c}{$0.91 \pm 0.26$}  & \multicolumn{2}{c}{$2.81 \pm 0.45$} &\\
 Mass ratio & \multicolumn{2}{c}{$0.22 \pm 0.06$} & \multicolumn{2}{c}{$0.15 \pm 0.02$} &
\enddata
\tablenotetext{a}{\citet{kirkpatrick10}}
\tablenotetext{b}{optical spectral type}
\tablenotetext{c}{NIR spectral type}
\tablenotetext{d}{Photometric relation from \citet{schlaufman10}}
\label{T_prop}
\end{deluxetable*}

\section{Discussion}
\subsection{Stability of the wide binaries}
\begin{figure}
\centering
\includegraphics[width=8.5cm,angle=0]{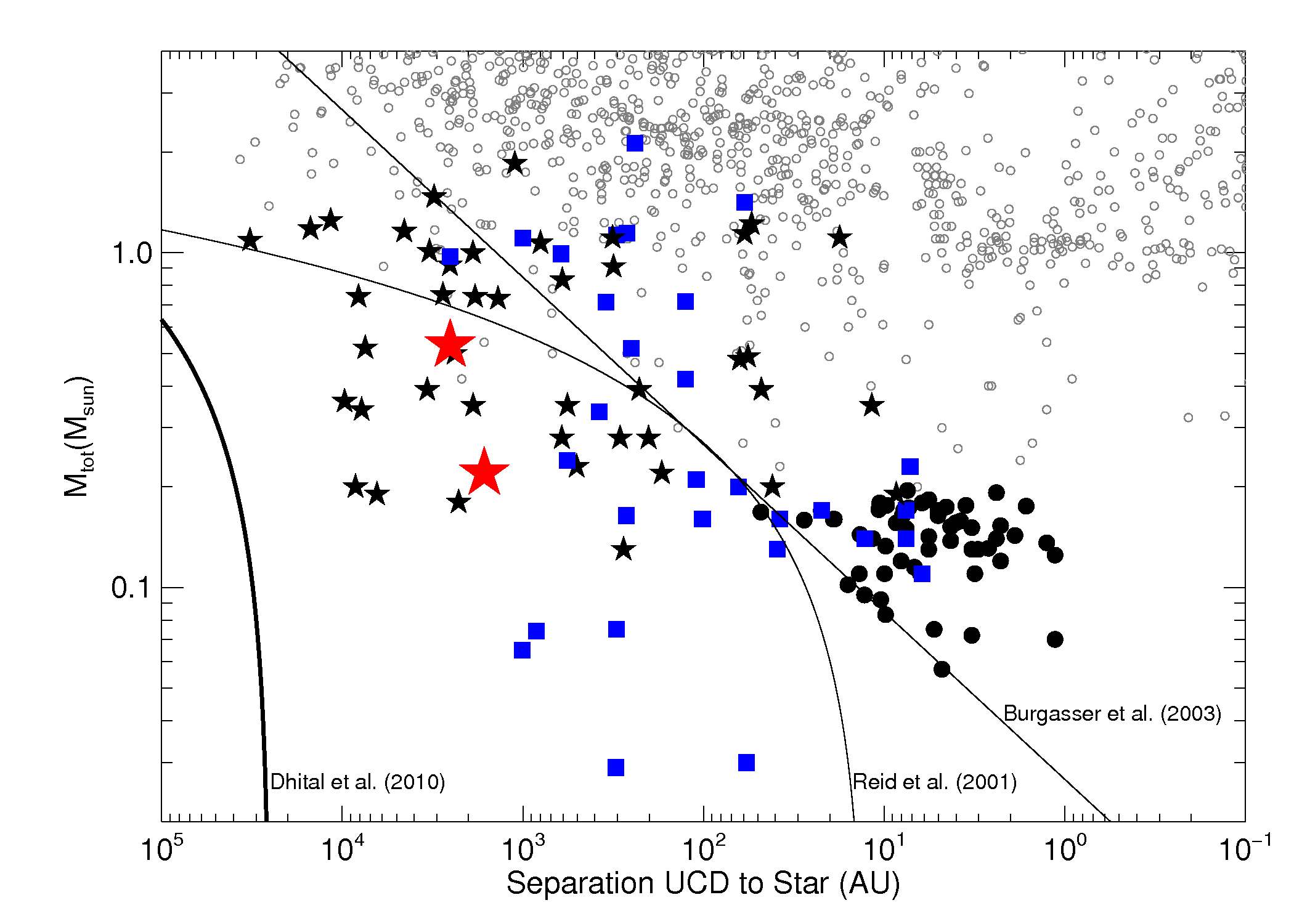}
\caption{A plot of the total mass vs. binary separation.  Objects marked with filled circles are tight low--mass systems. Wide systems (separation $>$100 AU) containing a UCD companion are marked as five point stars and those marked as blue squares are systems containing a tight or widely separated UCD with an age $<$ 500 Myr. Objects marked by open circles come from stellar companion catalogs. The new wide systems reported in this work are marked as red five-point stars.
We have over-plotted the \citet{reid01} curve (center) which distinguished the cut-off for the formation of wide stellar pairs as well as the \citet{burgasser03} line (specific for M$_{tot}<$0.2 M$_{\sun}$ field systems) and the log-normal limitation for systems with M$_{tot}$ $>$ 0.3 M$_{\sun}$ found in \citet{dhital10}. Data were taken from Figure 9 in \citet{faherty11}, see references therein for more details.}
\label{BE}
\end{figure}

Figure~\ref{BE} shows the total mass versus separation for a sample of binary systems.
We show the suggested empirical limits for the stability of VLM multiples from the studies of \citet{reid01} and \citet{burgasser03}, based on objects that were known at the time
of the respective works. Neither cutoff seems appropriate for
the collection of widely separated systems plotted (see
discussions in \citealt{faherty10} and \citealt{dhital10}). 
The log-normal limitation on the separation of binaries found by \citet{dhital10}
does encompass 2M0525-7425AB and all low-mass objects
in Figure~\ref{BE}.
\citet{zuckerman09} derived a cutoff for the binding energy of a system formed by
fragmentation as a function of the total system mass, assuming separations of 300~AU. 
Since a number of systems found in the field and young clusters are found at separations larger than 300 AU,
\citet{faherty10} used the Jeans length instead of the fiducial value of 300 AU to repeat the calculation
for two extreme mass ratio cases. 
Both 2M0525-7425AB and 2M1348-1344 are found above the \citet{zuckerman09} and \citet{faherty10} limits.

\citet{dhital10} calculate the average binary lifetimes, that generally depend on the total mass and separation
(see their Figure~19). With the parameters calculated in the previous sections, both systems reported here appear to be stable enough to survive longer than 10 Gyrs. 
Based on the component proper motions, excellent agreement in the independent distance estimates of each component, 
and stability arguments, we 
conclude that 2M0525-7425AB and 2M1348-1344AB are genuine binary systems.

\subsection{Mass-ratios}
\label{q}
Both wide binary systems reported in this work have relatively
low mass ratios($q=0.22$ for 2M1348-1344 and $q=0.15$ for 2M0525-7425). 
By analyzing a 
catalog of 1342 very-wide (projected separations of $\gtrsim 500\,$AU) low-mass (at least one mid-K to mid-M dwarf component) binaries, \citet{dhital10} find the distribution of mass ratios to be
strongly skewed toward equal-mass pairs: 85.5\% of pairs have masses within 50\% of each other. The
authors conclude that this does not result from observational biases, i.e. the survey 
is sensitive to pairs with much lower mass ratios. 
A similar result is obtained by \citep{reid97} in the volume-complete 8-pc sample, constituted mostly (80\%)
of M-dwarfs. However, the sensitivity to mass ratios in both works is reported for the entire samples, rather than for M dwarfs alone, 
and the sensitivity to low mass ratio systems will fall off with lower-mass primaries. 
In a study of multiplicity among M-dwarfs, \citet{fischer92} found that the distribution
of mass ratios is flat and perhaps declines at low mass ratios.
A number of low-$q$ binary systems containing ultracool 
components are known to date, some with the mass-ratios as low as 0.05 (e.g. \citealt{faherty10} and references therein).
A complete sample sensitive to ultracool dwarfs of the very lowest masses is needed to assess the distribution
of mass ratios in this regime.

\subsection{Higher-order multiplicity}

It has been found that binaries wider than $\sim$100~AU have a high rate of tertiaries, both in the stellar
and substellar regimes, and that the binary fraction of ultracool companions found in multiple systems appears to be significantly larger than that of field dwarfs \citep{burgasser05,tokovinin06,law10,faherty10}. 
This behavior has also been predicted by some dynamical models 
\citep{sterzik03,delgado04, whitworth06,kouwenhoven10}. Furthermore, dwarfs with spectral types of late L or early T
show a higher multiplicity fraction compared to other types (\citealt{liu06,burgasser06b}; but see also 
\citealt{goldman08}). This has been
attributed to the rapid evolution through the L/T transition phase, implying fewer sources per spectral subtype
for a given field sample \citep{burgasser07}.
Unresolved binaries can be identified by an apparent over-luminosity for a given spectral subtype, 
where an independent distance estimate is available (e.g. \citealt{dupuy12}), or by spectral deviations
from template spectra  (e.g. \citealt{burgasser10}).
\citet{burgasser10} identified several common trends that stand out when comparing the combined near-infrared spectra
of resolved L/T binaries with the template spectra of single objects. 

The SpeX spectrum of the L/T transition dwarf 2M0614+3950 (Figure \ref{2M0614}) 
is closely matched by the L9 spectral template, and 
does not show any of the deviations pointed out in \citet{burgasser10}.
For the two wide binaries, 2M0525-7425AB and 2M1348-1344AB, the distance estimates of the primaries 
are in excellent agreement with the independent distance estimates of the secondaries.
For example, by placing the T5.5 dwarf 2M1348-1344B at the distance of the primary augmented by its 1$\sigma$ uncertainty
+0.25 in its absolute J-magnitude, and -0.15 in its $J-H$ color, still consistent with the derived spectral type. 
Also, all of the component spectra presented here 
do not appear to deviate significantly from the spectral templates. We note that the classification of 2M0525-7425B 
originates from \citet{kirkpatrick10}, who do not show a spectrum in the paper, but neither report any peculiarities.
We conclude that, while we have no reason to suspect any unresolved binarity in the objects discussed here, 
high resolution imaging and/or radial velocity study is necessary to prove this claim. 

\subsection{Predictions from theory}

Among the multiple systems produced in hydrodynamical simulations of star cluster formation by \citet{bate09},
about $10\%$ are stars ($\geq0.1\,$M$_{\odot}$) with VLM ($<0.1\,$M$_{\odot}$) companions. 
About 30\% of these star/VLM binary systems have semi-major axes larger than 1000~AU, and several
objects with primary masses similar to those of the two systems reported here are created.  
Although there appears to be no statistically significant dependence of the frequency of star/VLM binaries on primary mass, 
the typical separation of such binaries seems to increase strongly with the primary mass. 
As for the mass ratios for the systems with 0.1 -- 0.5 M$_{\odot}$ primaries, \citet{bate09} predicts a slight
preference towards near-equal mass systems. However, a significant number of systems with low-mass ratios is produced
in the simulations, with about 15\% of the binaries having $q \leq 0.2$.
Another prediction of this model is a corrrelation between mass ratio and separation for the binaries, 
with closer systems having a preference for equal masses. The median mass ratio for binary separations in range 1000 -- 10$^4$ AU
is 0.17, comparable to the mass ratios observed in our two binary systems.
Simulations by \citet{stamatellos09} show that very wide (up to 10~000 AU), low-mass (secondary mass $<$ 80 M$_J$) companions can be formed in fragmenting stellar disks. In the simulation of a 0.7$\,$M$_{\odot}$ star with a disk of an equal
mass and a radius of 400~AU, wide (400-10~000~AU) binary companions to the central star are predominantly brown dwarfs.
An alternative mechanism to form wide binaries was proposed by \citet{kouwenhoven10} and \citet{moeckel11}.
Their N-body simulations of star clusters in dissolution show that binaries with separations larger than 1000~AU can be 
formed by dynamical capture of initially unbound objects. While the wide binary fraction from \citet{kouwenhoven10} varies
significantly with the initial conditions, by integrating over the initial cluster mass distribution they
predict a wide binary fraction of a few percent in the Galactic field. 
For wide systems with separations $>1000\,$AU and primary masses below $1.5\,$M$_{\odot}$, \citet{kouwenhoven10} models predict 
the increase in the number of the systems with decreasing the mass ratio. The mass-ratio distributions peaks around $q=0.1-0.3$.

The existence of binaries with total system masses, mass ratios and separations similar to the two systems reported here 
is clearly supported by predictions of different theoretical models. A proper statistical analysis of complete wide binary
samples is needed to asses their overall frequency and distribution of properties such as mass ratio, 
and to be able to compare these observational results with the predictions of the models.

\section{Summary and conclusions}
\label{summary}
We have presented the results of our search for common proper motion pairs containing
ultracool components, based on 2MASS and Preliminary WISE data catalogs, followed by optical or NIR 
spectroscopy of individual candidate components. We report the discovery of two wide binaries containing 
ultracool components, and one ultracool dwarf at the L/T transition.

2M1348-1344AB consists of a M4.5 primary with a T5.5 companion, at a distance $\sim20\,$pc from the Sun, and 
at separation of $~1400\,$AU. An M3 primary and its L2 companion comprise the system 2M0525-7425AB, found lying at
a distance $\sim45\,$pc, and separated by $~2000\,$AU. Based on the matching proper motions
and excellent agreement in distance estimates of the components, as well as on the fulfilled stability
criteria, we conclude that both systems are genuine low-mass binaries. Both primaries are found to have
metallicities close to that of our Sun, and lack H$\alpha$ emission, allowing us to place
the lower limits on the ages of the two systems at 1.5 Gyr (2M0525-7425) and 4 Gyr (2M1348-1344). 
We also report the discovery of 2M0614+3950, an ultracool dwarf with spectral type L9, lying at a distance
$\sim 26\,$pc. 

The three objects reported here add to the compendium of ultracool objects discovered by WISE, which is certain 
to grow with the recent survey's all sky release. Further characterization of the L/T transition dwarf and the components 
of the two new binary systems will provide stringent tests for understanding of ultracool atmospheres, as well as 
for the formation models of BDs and BD binaries. 2M1348-1344AB is particularly important in this sense, 
being one of a handful of known late-T very wide companions to low mass stars.

\acknowledgments
We thank David Lafreni\`ere for his
contribution to writing software used for this project.
The research was supported in part by grants from
the Natural Sciences and Engineering Research Council
(NSERC) of Canada to RJ.
RJ's work was also supported in part by the Radcliffe Institute for
Advanced Study at Harvard University.
RK acknowledges support from Cento de Astrof\'isica de Valpara\'iso and DIPUV 23/2009.
SM's work is supported by NASA through grant 10-ADAP10-0130.
This work was co-funded under the Marie Curie Actions of the European
Commission (FP7-COFUND).
JB is supported by the Chilean Ministry for the Economy,
Development, and Tourism's Programa Iniciativa Cient\'{i}fica Milenio through grant P07-021-F,  awarded to The Milky
Way Millennium Nucleus and  FONDECYT Reg. No. 1120601.
This publication makes use of data products from the Two Micron All
Sky Survey, which is a joint project of the University of Massachusetts
and the Infrared Processing and Analysis Center/California Institute of
Technology, funded by the National Aeronautics and Space Administration
and the National Science Foundation. This research has benefitted from 
the SpeX Prism Spectral Libraries, maintained by Adam Burgasser 
at 
{\tt http://pono.ucsd.edu/\textasciitilde adam/browndwarfs/spexprism/}.
This research has also benefited from the M-dwarf spectra available from {\tt http://www.dwarfarchives.org}.


\clearpage

\end{document}